# The suppression of superconductivity in MgCNi$_3$ by Ni-site doping

M. A. Hayward[a*], M. K. Haas[a], A. P. Ramirez[b], T. He[a], K.A. Regan[a], N. Rogado[a], K. Inumaru[a], R. J. Cava[a]

[a]*Department of Chemistry and Princeton Materials Institute, Princeton University, Princeton NJ*

[b] *Condensed Matter and Thermal Physics Group, Los Alamos National Laboratory, Los Alamos NM*

*Corresponding author: Tel : 609-258-5556, Fax: 609-258-6746,

E-mail: mhayward@princeton.edu

## Abstract

The effects of partial substitution of Cu and Co for Ni in the intermetallic perovskite superconductor MgCNi$_3$ are reported. Calculation of the expected electronic density of states suggests that electron (Cu) and hole (Co) doping should have different effects. For MgCNi$_{3-x}$Cu$_x$, solubility of Cu is limited to approximately 3% (x = 0.1), and T$_c$ decreases systematically from 7K to 6K. For MgCNi$_{3-x}$Co$_x$, solubility of Co is much more extensive, but bulk superconductivity disappears for Co doping of 1% (x = 0.03). No signature of long range magnetic ordering is observed in the magnetic susceptibility of the Co doped material.

The recent discovery of superconductivity in the intermetallic perovskite $MgCNi_3$ (1) has provided a link between two major families of superconducting materials, namely intermetallic compounds and perovskite based oxides. The observation of superconductivity in $MgCNi_3$ is surprising given the large amount of nickel present, an element usually associated with magnetism. Superconductivity in nickel rich intermetallics is not unprecedented, however. The quaternary borocarbides, $RENi_2B_2C$, are prime examples (2-5); $MgCNi_3$ can be considered to be a three dimensional analog of these layered materials. Given the presence of strong antiferromagnetic fluctuations and the suggestion of unconventional pairing in the superconducting state of $LuNi_2B_2C$ (6,7), a more detailed study of the properties of $MgCNi_3$ is of interest. Here we report the effects of Ni-site doping on the superconductivity of $MgCNi_3$. The dopants selected, Co and Cu, were motivated by electronic density of states calculations, also reported here. Qualitatively different behavior is observed for the two cases: the Cu doping suppresses $T_c$ in a manner consistent with expectations based on simple band filling. Conversely, the much stronger suppression of superconductivity for Co doping is opposite to a simple band filling effect, suggesting an important role of spin fluctuations in pair breaking

Our chemical doping experiments were motivated by the unusual band structure of this material. The electronic band structure was calculated using the Wein97 package (8). This program employs a scalar relativistic full-potential linear augmented plane-wave (FLAPW) method, with Correlation and exchange effects treated using the generalized gradient approximation (GGA) of Purdue, Burke and Ernzerhof (9) within density-functional theory (10). The Bloechl et. al. tetrahedron method (11) was used to calculate density of states (DOS), with a total of 56 irreducible k-points from a grid of $10 \times 10 \times 10$ reducible k-points

Figure 1 plots the calculated total electronic density of states (DOS) against energy for MgCNi$_3$. The most striking feature of this plot is a very large, narrow energy peak in the DOS just below the Fermi energy (E$_F$). E$_F$ is on the high-energy side of the peak, at an electron count approximately 0.5 electrons per formula unit higher than the peak maximum. Analysis of the individual contributions to this peak indicates that it is 80 – 90 % Ni d in character. This type of narrow energy peak is a characteristic typical of materials that display strong magnetic interactions. It suggests that MgCNi$_3$ is on the border of magnetic instability, and that hole doping should induce a transition from superconductivity to magnetism. Electron doping, on the other hand, should lead to a decrease in T$_c$ simply due to a decreasing density of states.

In order to investigate the doping effects suggested by the electronic structure calculations, 0.5g samples with compositions MgCNi$_{3-x}$M$_x$ (M = Cu x = 0.03, 0.06, 0.09; M = Co x = 0.03, 0.06, 0.09, 0.15, 0.20, 0.25, 0.5, 0.75) were prepared. Starting materials were bright Mg flakes (Aldrich) fine Ni powder (Alfa Aesar), fine Co powder (Johnson Matthey) and carbon black (Alfa Aesar). Previous work on MgCNi$_3$ (1) indicated that it is necessary to prepare samples with a stoichiometric excess of both magnesium and carbon in order to obtain optimal carbon content. All samples were therefore prepared with the initial stoichiometry Mg$_{1.2}$C$_{1.5}$Ni$_{3-x}$M$_x$. The well mixed starting materials were pressed into pellets and placed on Ta foil, which in turn was placed on an Al$_2$O$_3$ boat and fired in a quartz tube furnace under a flowing atmosphere of 95% Ar, 5% H$_2$. Samples were heated for half and hour at 600°C followed by one hour at 900°C. After cooling samples were ground, pressed into pellets, and heated for an additional hour at 900°C. Powder X-ray diffraction confirmed that all samples reacted to form a single cubic phase ($a \cong 3.81$ angstroms) with no observable impurities. (The Mg excess volatizes during the synthesis and the expected graphite second phase can be observed by neutron diffraction in materials prepared in this manner (1).

The superconducting transitions of the doped materials were characterized by zero-field cooled DC magnetization measurements performed on loose powder samples in an applied DC field of 15 Oe. Figure 2 shows the magnetization data collected from the x = 0, 0.03, 0.06 and 0.09 copper and cobalt doped samples over a relatively wide temperature range and Figure 3 shows the region around $T_c$. The difference between doping with copper or cobalt on the evolution of magnetic behavior is striking. Doping $MgCNi_3$ with low concentrations of copper (electron doping) progressively lowers the superconducting transition temperature from $T_c \cong 7K$ (x = 0) to $T_c \cong 6K$ (x = 0.09). This suppression of $T_c$ can be readily attributed to the expected decrease in the DOS at $E_F$ with increasing electron count predicted by the electronic band structure calculations illustrated in Figure 1.

Cobalt doping has a qualitatively very different effect. The addition of small amounts of cobalt, even as little as 1%, rapidly quenches bulk superconductivity. This is seen in Figure 3 by a dramatic decrease in the amount of diamagnetism with no change in the diamagnetic on-set temperature. The observation of any diamagnetism at all is most likely due to an unavoidable but small amount of inhomogeneity in the Co distribution, leaving some small regions of the sample Co free and superconducting with the undoped $T_c$. This is contrary to the conventional expectation that $T_c$ should rise with the increase in DOS at $E_F$, predicted by Figure 1, that should occur upon hole doping. For comparison, previous work studying the effects of doping the two dimensional materials, $YNi_2B_2C$ (12) and $LuNi_2B_2C$ (13,14) report a drop in the superconducting transition temperature on the substitution of Cu or Co for Ni. Such behavior is consistent with a simple band filling for these compounds since both electron and hole doping are expected to reduce the DOS at $E_F$ (15,16). In addition single crystal magnetization measurements performed on $LuNi_{2-x}Co_xB_2C$ have explicitly ruled out a pair breaking mechanism for the observed suppression of $T_c$. (14)

The observed rapid loss of superconductivity, on the substitution of Co for Ni, is more consistent with the magnetic quenching of superconductivity. This suggests that Co is a very strongly acting magnetic impurity rather than a source of hole doping.

In order to elucidate the magnetic consequences of the Co and Cu doping on $MgCNi_3$, the normal state susceptibilities of these materials were measured between 300 and 10K. We observed a significant low-field magnetization, presumably due to the presence of very small amounts (less than 1%) of unreacted, ferromagnetic Ni metal in all preparations. Therefore, in order to obtain the intrinsic susceptibility, we first determined a linear magnetization versus field range and then measured the magnetization at two field values in this range. The intrinsic susceptibility of each compound was thus obtained from the difference in M between applied fields of 4T and 2T at each measurement temperature.

The normal state magnetic susceptibilities of $MgCNi_3$ and the doped variants obtained in this manner are shown in Figure 4. The undoped material shows a normal state susceptibility of $1.7 \times 10^{-4}$ emu/moleNi compared to the observed $\gamma \sim 10$

mJ/moleNiK$^2$ (1). This yields a Wilson ratio of 1.2 indicating an enhancement of the susceptibility due to spin fluctuations. The Cu doping is seen to suppress the normal state susceptibility, fully consistent with the expected decrease in γ suggested by the band structure calculation. The data in Figure 4 show that doping MgCNi$_3$ with cobalt does not result in bulk ferromagnetism at any doping level up to x = 0.75. The Co doping levels employed are sufficient to pass through the calculated peak in the electronic density of states, and therefore the naïve expectation that ferromagnetism might occur at such a high narrow DOS peak is not fulfilled. In addition, the normal-state susceptibility data do not scale straightforwardly as a function of cobalt concentration, which implies picturing cobalt as a simple localized magnetic impurity center is inappropriate.

Figure 5 shows a plot of the magnetic susceptibility at 100K and the calculated density of electronic states as a function of electron count for the composition region studied. It can be seen that the magnetic susceptibility and the calculated density of states roughly vary in the same manner, with electron count. This suggests that the primary effect of Co doping is to change the density of states in a manner consistent with the electronic structure calculation. This increase in DOS does not however lead to an increase in T$_c$. Why the Co doping appears to act in a manner consistent with the expected density of states and yet immediately quenches the superconductivity is enigmatic.

In summary, we observe that Ni-site substitution leads to a suppression of superconducting behavior in MgCNi$_3$. Cu doping results in the depression of the superconducting critical temperature in a manner a kin to that observed in the 2-dimensional borocarbides, YNi$_2$B$_2$C and LuNi$_2$B$_2$C. Substitution of Ni with Co rapidly quenches superconductivity, contrary to the expectations of simple band filling, by what is an as yet unresolved magnetic interaction. There is no evidence for bulk ferromagnetism in any of the Co doped samples. The difference in the two types of behavior is therefore generally consistent with what is expected from the band structure, but not in a trivial manner.

**Acknowledgement**

The work at Princeton was supported by the National Science Foundation Division of Materials Research and the Department of Energy, Division of Basic Energy Sciences.

**Figures**

Figure 1: The calculated energy dependence of the electronic density of states for MgCNi$_3$. E$_F$ = Fermi energy. Inset: detail of region in vicinity of E$_F$.

Figure 2: Magnetic characterization of the superconducting transitions for MgCNi$_{3-x}$Cu$_x$ (upper panel) and MgCNi$_{3-x}$Co$_x$ (lower panel).

Figure 3: Enlargement of the magnetic characterization of the superconducting transitions for MgCNi$_{3-x}$Cu$_x$ (upper panel) and MgCNi$_{3-x}$Co$_x$ (lower panel), illustrating the contrasting behavior on doping with Cu vs Co.

Figure 4: Normal state magnetic susceptibilities of MgCNi$_{3-x}$Cu$_x$ and MgCNi$_{3-x}$Co$_x$.

Figure 5: The magnetic susceptibility at 100K and calculated electronic density of states plotted as a function of electron count for the composition range studied.

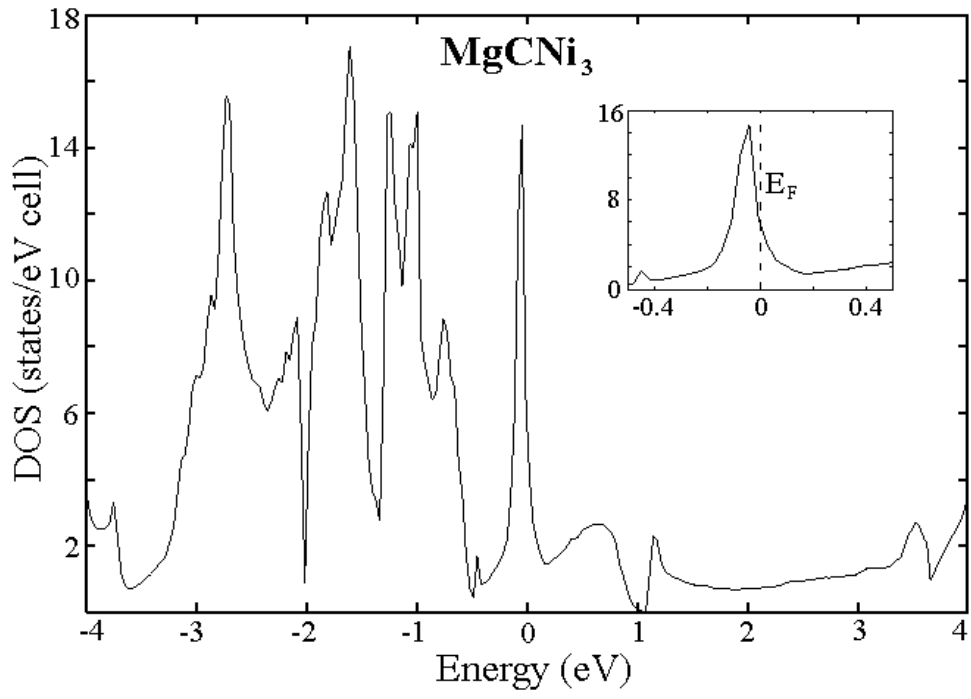

Figure 1

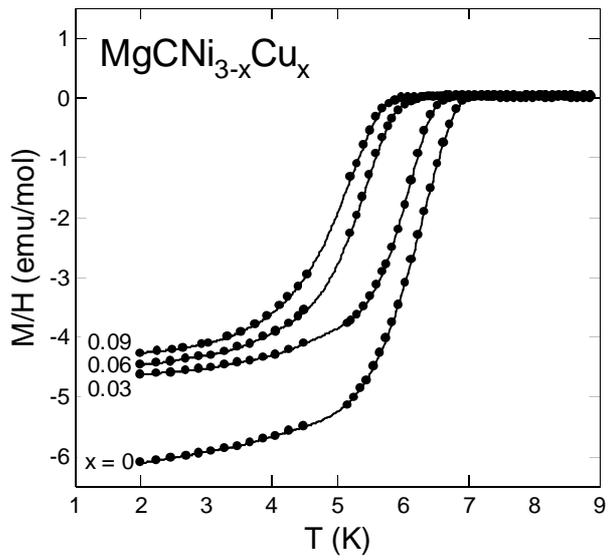

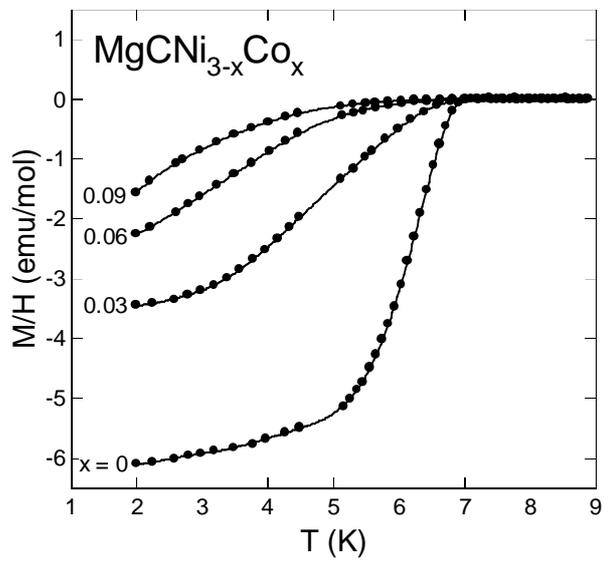

Figure 2

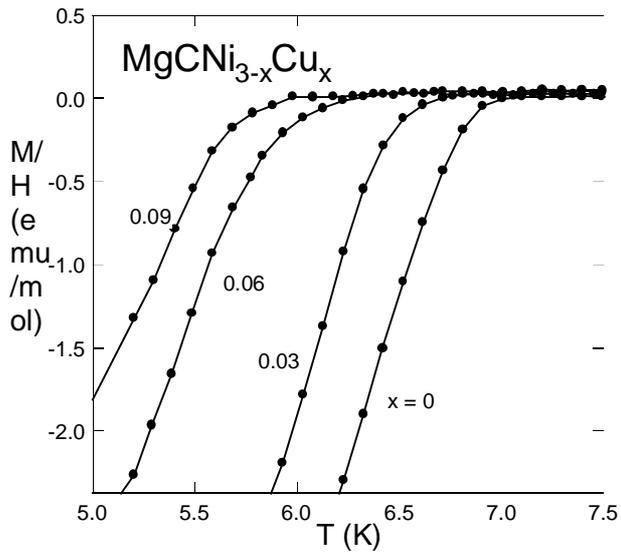

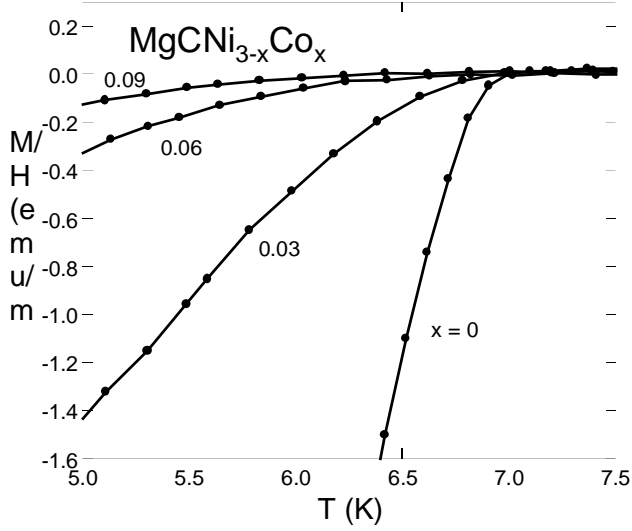

Figure 3

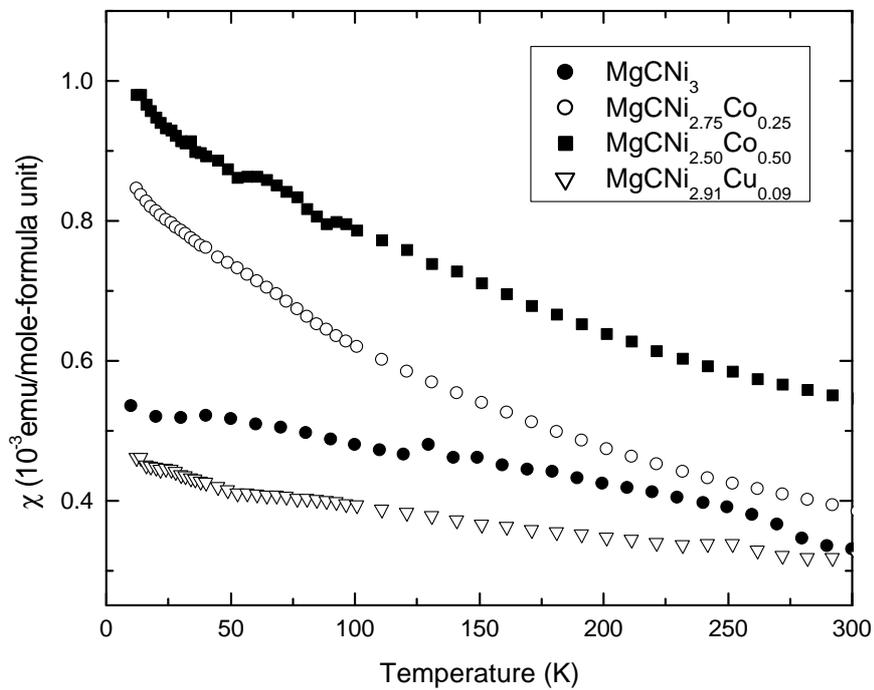

Figure 4

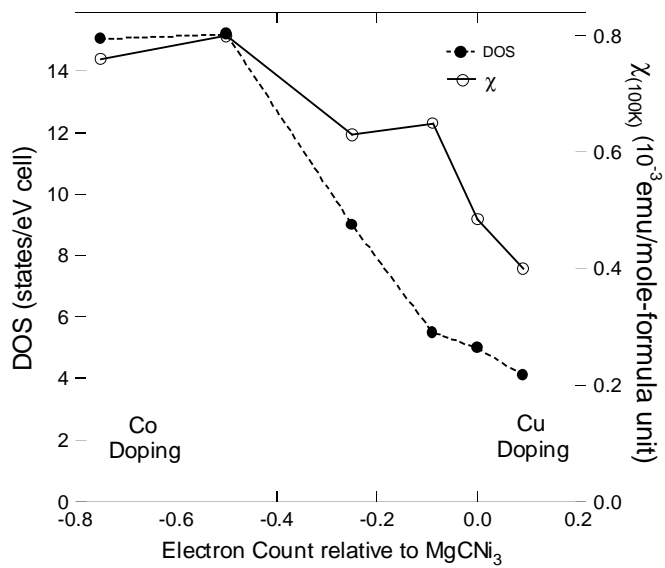

Figure 5